\documentclass{mn2e}
\usepackage{epsfig}
\usepackage{subfigure}
\usepackage{multirow}
\usepackage{amssymb}
\usepackage[figuresright]{rotating}

\newcommand{\msun}{$\mathrm{M_{\sun}}$}

\newcommand{\kms}{$\mathrm{km \, s^{-1}}$}

\title[Double-core evolution and the formation of neutron-star
      binaries]{Double-core evolution and the formation of
      neutron-star binaries with compact companions}

\author[Dewi, Podsiadlowski {\rm\&} Sena]
       {J. D. M. Dewi$^{1,2}$\thanks{email: jasinta@ast.cam.ac.uk (JDMD), 
                                            podsi@astro.ox.ac.uk (PhP)}, 
        Ph. Podsiadlowski$^1$\footnotemark[1], A. Sena$^1$\\
        $^1$Department of Astrophysics, University of Oxford, 
            Keble Road, Oxford OX1 3RH, UK\\
        $^2$Institute of Astronomy, University of Cambridge,
            Madingley Road, Cambridge CB3 0HA, UK}
\date{Accepted . Received ; in original form }

\pagerange{\pageref{firstpage}--\pageref{lastpage}}
\pubyear{}

\begin{document}

\maketitle

\label{firstpage}

\begin{abstract}
We present the results of a systematic exploration of an alternative evolutionary scenario to form double neutron-star binaries, first proposed by Brown (1995), which does not involve a neutron star passing through a common envelope.  In this scenario, the initial binary components have very similar masses, and both components have left the main sequence before they evolve into contact; preferably the primary has already developed a CO core. We have performed population synthesis simulations to study the formation of double neutron star binaries via this channel and to predict the orbital properties and system velocities of such systems. We obtain a merger rate for DNSs in this channel in the range of 0.1 -- 12~$\mathrm{Myr}^{-1}$.  These rates are still subject to substantial uncertainties such as the modelling of the contact phase.
\end{abstract}


\section{Introduction}
\label{doublecore:sec:intro}

Double neutron star (DNS) systems, i.e. binaries consisting of a radio pulsar and a usually undetected second neutron star, have provided an excellent test bed for general relativity theory. The orbital decay observed in DNSs is the first indirect evidence for gravitational-wave radiation. The merger of the two neutron stars in DNSs is expected to be a major source for gravitational-wave detection by the Laser Interferometer Gravitational Wave Observatory (LIGO), Virgo, or the Laser Interferometer Space Antenna (LISA). Such mergers are also thought to be potential progenitors of the short-duration Gamma-Ray Bursts (GRBs). The most commonly discussed formation channel of DNSs, which we refer to as the standard scenario (e.g. Bhattacharya \& van den Heuvel 1991), involves a high-mass X-ray binary (HMXB) phase at an intermediate stage. The HMXB eventually undergoes Roche-lobe overflow (RLOF) which leads to a common-envelope (CE) phase. The expected outcome of this phase is a binary consisting of the helium core of the massive star and an X-ray emitting neutron star.

Chevalier (1993) argued that a neutron star which moves in the dense environment of a stellar envelope can accrete a considerable amount of matter, which may push the neutron star above the maximum mass allowed for neutron-star matter and lead to the formation of a black hole. Because of this argument, Brown (1995) suggested that HMXBs passing through a CE generally produce helium-star/black-hole binaries rather than helium-star/neutron-star binaries -- and hence do not lead to the formation of DNSs in this standard scenario. To avoid the problem of hypercritical accretion onto a neutron star, Brown (1995) (see also Wettig \& Brown 1996; Bethe \& Brown 1998) suggested that DNSs originate from binaries consisting of two helium stars in a close orbit, which we will refer to as the double core CE scenario. The first-born neutron star accretes part of the stellar wind from the other helium star which may weaken its magnetic field strength and spin up the neutron star. An important constraint for the formation of a double helium-star binary in a close orbit is that the systems start with two components of almost equal mass (within 4 per cent).

However, strong accretion might not have to occur if the companion is extended (Chevalier 1993), if rotation of the magnetized neutron star is taken into account (Chevalier 1996) or if there is a strong outflow from a disc around the neutron star (Armitage \& Livio 2000). Furthermore, Taam, King \& Ritter (2000) suggested that, considering the low rates of accretion characteristic of the low density envelopes of more evolved stars in combination with the short duration of the CE phase, accretion may not be important. These counter-arguments suggest that the formation of DNSs via the standard channel cannot be ruled out at this stage. Moreover, there are good arguments suggesting that at least for some pulsars in binaries the neutron star must have passed through a common envelope (e.g. Tauris, van den Heuvel \& Savonije 2000).

We will not argue in this paper whether or not a neutron star in a CE phase will become a black hole. Nevertheless, the standard scenario clearly is not valid for the situation where the initial binary has almost equal mass components, and hence it is interesting to investigate whether binary systems with these particular initial conditions could also lead to the formation of DNSs, which is the main purpose of this paper. We refer to this channel with stars of almost equal initial masses as the double-core scenario. Using binary population synthesis simulations, we study the expected observational properties as well as the birth and merger rates of DNSs formed via the double-core scenario and compare them to the results of the standard scenario.

We describe the detail of the double-core scenario, as well as the basic assumptions and the code used in this work, in Sect.~\ref{doublecore:sec:method}. The results are presented in Sect.~\ref{doublecore:sec:result} and discussed further in Sect.~\ref{doublecore:sec:conclusion}.

\section{The double-core scenario}
\label{doublecore:sec:method}

\subsection{The evolutionary scenario}
\label{doublecore:subsec:scenario}

The double-core scenario we consider in this paper is a slight variation of the original proposal by Brown (1995), i.e.\ it starts with a binary consisting of two stars of almost equal initial masses. The main difference to the original scenario is that we put more constraints on the initial masses and periods for the following reason. The pulsars in DNSs, which are the first-born neutron stars, are recycled, and this recycling process is assumed to be due to mass accretion from the progenitors of the pulsar companions. Therefore, it is necessary that the first neutron star has sufficient time to accrete matter from the companion.

Binary calculations to model the formation of contact binaries have been presented by, e.g., Pols (1994), Wellstein, Langer \& Braun (2001) and Nelson \& Eggleton (2001). However, evolution through the contact phase has not been modelled (Webbink 2003). Here we assume that the outcome of a contact phase is a close binary consisting of the cores of the primary and the secondary star. If the masses of the components are very similar, i.e.\ differ by less than 0.3 per cent (Podsiadlowski, Joss \& Hsu 1992), the outcome of the contact phase is that the two cores are more or less in the same evolutionary phase, i.e.\ produces He + He or CO + CO binaries. In this case, the two cores also have almost equal masses and collapse into neutron stars with a relatively short time between the two collapse phases. In this situation, the first-born neutron star has a very limited possibility to accrete matter from the companion. Moreover, the probability of having a binary system with such similar initial masses is extremely low, and therefore we would not expect to form many recycled DNSs binaries in this case. A channel to form non-recycled DNSs from almost-twin binaries has previously been studied by Belczy\'{n}ski \& Kalogera (2001).

\begin{figure}
  	 \centerline{\includegraphics[width=70mm]{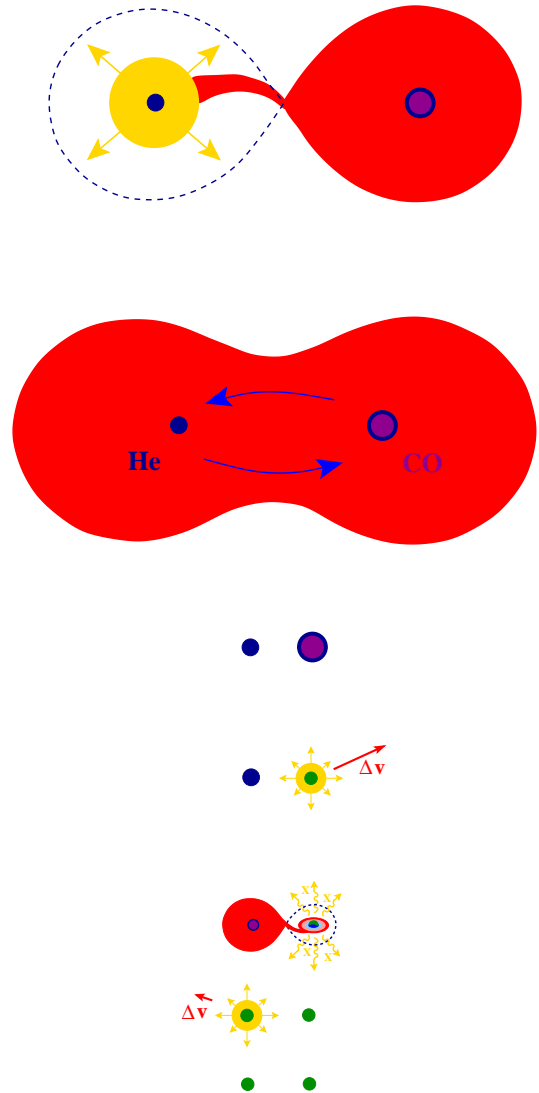}}
	 \caption[]{The formation of a double neutron star binary in
	 the double-core scenario. The binary consists initially of
	 two stars of almost equal mass (typically within 3 to 7 per
	 cent), and the primary fills its Roche lobe when is has
	 already developed a CO core while the secondary has already
	 left the main sequence. The system experiences dynamical mass
	 transfer, leading to the formation of a common-envelope (CE)
	 and a spiral-in phase. After the ejection of the CE, the
	 system is a close binary consisting of a CO and a He
	 star. The CO star soon collapses to form the first neutron
	 star, producing a helium-star/neutron-star binary. The helium
	 star may transfer matter to the neutron star before it
	 eventually explodes to produce the second neutron star.}
	 \label{doublecore:fig:scenario}
	\end{figure}

To avoid having two cores of almost equal masses and to allow the first neutron star to accrete matter, we propose that at the onset of the contact phase, the primary is in a more advanced stage of evolution than the secondary. This can be achieved, e.g., if the primary has developed a CO core when the secondary starts core helium burning, or if the primary has started core helium burning while the secondary is at the end of main sequence or at the beginning of shell hydrogen burning. In this paper we will consider the first option, i.e.\ the outcome of the contact phase is the CO core of the primary and the helium core of the secondary. The CO core collapses into a neutron star, which then accretes matter from the wind and/or the transferred matter during RLOF from the helium star. The latest stage of this scenario is the same as in the standard scenario, i.e.\ a helium-star/neutron-star (HeS-NS) binary in a close orbit. This formation of DNS through the double-core scenario is presented in Fig.~\ref{doublecore:fig:scenario}.

\subsection{Methods of calculations}
\label{doublecore:subsec:code}

We performed Monte Carlo simulations of binary systems with almost equal mass components. The initial primary mass distribution was assumed to follow a Salpeter mass function.  Since our main aim is to investigate the formation of DNSs, we therefore limited the initial primary masses to be $8 < M_{1}/{\mathrm{M_{\sun}}} < 25$, i.e.\ the presently favoured mass range in which single stars leave neutron stars as remnants. The secondary mass is chosen such that the mass ratio of the system, $q = M_{2}/M_{1}$, follows a uniform distribution. Because our investigation is on almost equal mass binaries, only systems with $q > q_{\mathrm{crit}}$ are taken into account, where the critical mass ratio $q_{\mathrm{crit}}$ is determined from single star models so that the secondary will already have finished its main-sequence phase when the primary has finished its helium-core burning phase. The value of $q_{\rm crit}$ is therefore a function of $M_1$, varying from 0.97 for $M_1= 8\,{\mathrm{M_{\sun}}}$ to 0.93 for $M_1= 25\,{\mathrm{M_{\sun}}}$. For the distribution of orbital periods we take a uniform distribution in $\log~P$, with $10^{-1} < P\mathrm{(d)} < 10^{4}$.

To obtain the CO core of the primary and the He core of the secondary at the end of the contact phase, we limit our simulations to binary systems in which the primary undergoes case C mass transfer, i.e.\ its Roche radius follows $R_{\mathrm{C-min}} < R_{\mathrm{L,1}} < R_{\mathrm{C-max}}$, where $R_{\mathrm{C-min}}$ and $R_{\mathrm{C-max}}$ are the minimum and maximum radius for case C mass transfer from the primary. The minimum radius for case C mass transfer depends on how the star evolves accross the Hertzsprung-Russell (H-R) diagram, in particular on whether it spends most of its helium core-burning phase as a blue supergiant or whether it first experiences an extended helium core-burning phase as a red supergiant (see, e.g., Podsiadlowski et al.\ 1992). In the latter case, the minimum radius for case C mass transfer would be substantially larger, leading to fewer systems that can experience case C mass transfer. The details of the radius evolution of massive stars is still a major unresolved problem in the theory of evolution of massive stars; it depends on the treatment of convection, including the effects of convective overshooting and semi-convection, the treatment of mass loss and other mixing processes (for further discussion see, e.g., Langer \& Maeder 1995). In order to be able to assess the importance of these uncertainties, we consider two extreme options, where in one we assume that all stars first experience helium core-burning as red supergiants, leading to a high minimum radius for case C mass transfer, and where in the other we assume that the star spends most of its helium core-burning phase as a blue supergiant, leading to a low minimum radius for case C mass transfer.  The corresponding radii, $R_{\mathrm{C-min}}$ and $R_{\mathrm{C-max}}$ were estimated from single-star models and their radii in the respective phases.

The separation after the contact phase, $a_{\mathrm{f}}$, is determined as in the case of the standard single-core common-envelope evolution (Webbink 1984), i.e.\ the envelopes are ejected on the expense of the change in the orbital energy, $E_{\mathrm{env}} = \eta \, \Delta E_{\mathrm{orb}}$, where the envelope binding energy is
   \begin{eqnarray}
   E_{\mathrm{env}} = - \frac{G M_1 M_{\mathrm{e,1}}}
                             {\lambda_1 R_{1}}
                      - \frac{G M_2 M_{\mathrm{e,2}}}
                             {\lambda_2 R_{2}},
   \label{doublecore:eq:binding}
   \end{eqnarray}
and the change in orbital energy is
   \begin{eqnarray}
   \Delta E_{\mathrm{orb}} = - \frac{G M_{\mathrm{c,1}} M_{\mathrm{c,2}}}
                                    {2 a_{\mathrm{f}}}
                             + \frac{G M_1 M_2}
                                    {2 a_{\mathrm{i}}}.
   \label{doublecore:eq:orbital}
   \end{eqnarray}
At the onset of mass transfer, $R_{1} \sim R_{\mathrm{L,1}}$. The secondary is within its Roche lobe but expands as soon as it accretes matter from the primary, such that at the onset of the contact phase $R_{2} \sim R_{\mathrm{L,2}}$. Prior to the mass transfer phase, we neglect the stellar wind mass loss such that the Roche radii do not change. Using eqs.~\ref{doublecore:eq:binding} and \ref{doublecore:eq:orbital}, the reduction in separation can be expressed as
   \begin{eqnarray}
   \frac{a_{\mathrm{f}}}{a_{\mathrm{i}}} = 
   \frac{0.5 \eta M_{\mathrm{c,1}} M_{\mathrm{c,2}}}
        {0.5 \eta M_1 M_2 + 
         M_1 M_{\mathrm{e,1}}/\lambda_1 r_{\mathrm{L,1}} + 
         M_2 M_{\mathrm{e,2}}/\lambda_2 r_{\mathrm{L,2}}}.
   \label{doublecore:eq:separation}
   \end{eqnarray}
Here $\eta$ is an efficiency parameter for the envelope ejection, taken to be unity. The masses $M_{j}$, $M_{\mathrm{c},j}$ and $M_{\mathrm{e},j}$ ($j = 1, 2$ for primary and secondary, respectively) are the stellar mass, the core mass and the envelope mass. The CO and He core masses were taken from standard single star models for solar metallicity and adopting the Schwarzschild criterion for convection and included 0.25 pressure scale heights of convective overshooting.  $r_{\mathrm{L},j} = R_{\mathrm{L},j} / a_{\mathrm{i}}$, where $R_{\mathrm{L},j}$ is the Roche-lobe radius, obtained as in Eggleton (1983), and $a_{\mathrm{i}}$ is the separation at the onset of the contact phase. $\lambda_{j}$ is the envelope binding energy parameter, which depends on the stellar mass and evolutionary stage (Dewi \& Tauris 2000; Tauris \& Dewi 2001). For the purpose of this work we take a constant value of $\lambda$, i.e.\ 0.1 -- which is the average $\lambda$ value for high-mass stars (Dewi \& Tauris 2001). The exact value of $\lambda$ derived from the whole binding energy can be as low as 0.02 in stars more massive than $\sim$ 20~\msun\ and as high as 0.7 in 10~\msun\ on the giant branch. As alternative option, we also perform simulations with $\lambda = 0.5$, a value commonly adopted in population synthesis studies involving CE evolution. For each simulation we apply the same $\lambda$ value for both primary and secondary stars.

After the contact phase, we assume that there is no further episode of mass transfer from the CO core to the helium core. The CO core explodes in an asymmetric supernova (SN), in which a natal kick is imparted at the birth of the neutron star. We apply a Maxwellian kick velocity distribution with a dispersion $\sigma =$ 190~\kms\ (Model~HP; Hansen \& Phinney 1997) and a single Gaussian distribution with $\sigma =$ 290~\kms\ (Model~ACC; Arzoumanian, Chernoff \& Cordes 2002). The latter distribution is similar to the distribution found in the most recent study by Hobbs et al.\ (2005) who derived a kick velocity distribution with $\sigma =$ 265~\kms. The orbital period of the post-SN binary, as well as the system velocity after the explosion, are determined as in Brandt \& Podsiadlowski (1995). The explosion is assumed to leave a 1.4~\msun\ neutron star.

If the binary survives the SN explosion, it continues its evolution as a HeS-NS system. Extensive studies on mass transfer phase in HeS-NS binaries have been carried out by Dewi et al. (2002) and Dewi \& Pols (2003), taken into account a wide parameter space in helium star masses and orbital periods. During the mass-transfer phase, it was assumed that the neutron star accretes matter up to its Eddington limit and the rest of the transferred mass is lost to the system, carrying the specific angular momentum of the neutron star. We apply the results of the systematic studies by Dewi et al. (2002) and Dewi \& Pols (2003) to determine the orbital parameters of the system at the end of RLOF, which are the same as those prior to the explosion of the second SN. We apply the same kick velocity distribution as in the first SN. We also consider the recent suggestion that the second neutron star experiences a much lower kick velocity (Dewi, Podsiadlowski \& Pols 2005), i.e.\ by combining the above-mentioned kick distributions for the first neutron star and a distribution with $\sigma =$ 20~\kms\ for the second. These models are denoted by asterisks in Table~\ref{doublecore:tab:dns} (Models HP* and ACC*). To calculate the birth rate of DNS, we assume a SN rate of 0.01~$\mathrm{yr}^{-1}$. 

After a DNS is formed, it emits gravitational-wave radiation causing the decay of the orbit. We calculate the merger timescale of each system due to the emission of gravitational waves by applying the formula in Peters (1964). If a system merges in less than the Hubble timescale, which we assume to be 10~Gyr, it contributes to the total merger rate of DNSs.

\section{Results}
\label{doublecore:sec:result}

\subsection{Double neutron stars}
\label{doublecore:subsec:dns}

In Table~\ref{doublecore:tab:dns} we present the birth and merger rates of DNSs for each simulation. With lower $\lambda$ values ($\lambda = 0.1$), we obtain systems with smaller post-contact separations than applying $\lambda = 0.5$, and hence we would expect fewer systems to survive the contact phase. In general, applying a lower minimum radius for case C mass transfer means allowing for more systems to be included as DNS progenitors and hence we obtain higher birth and merger rates.

        \begin{figure*}
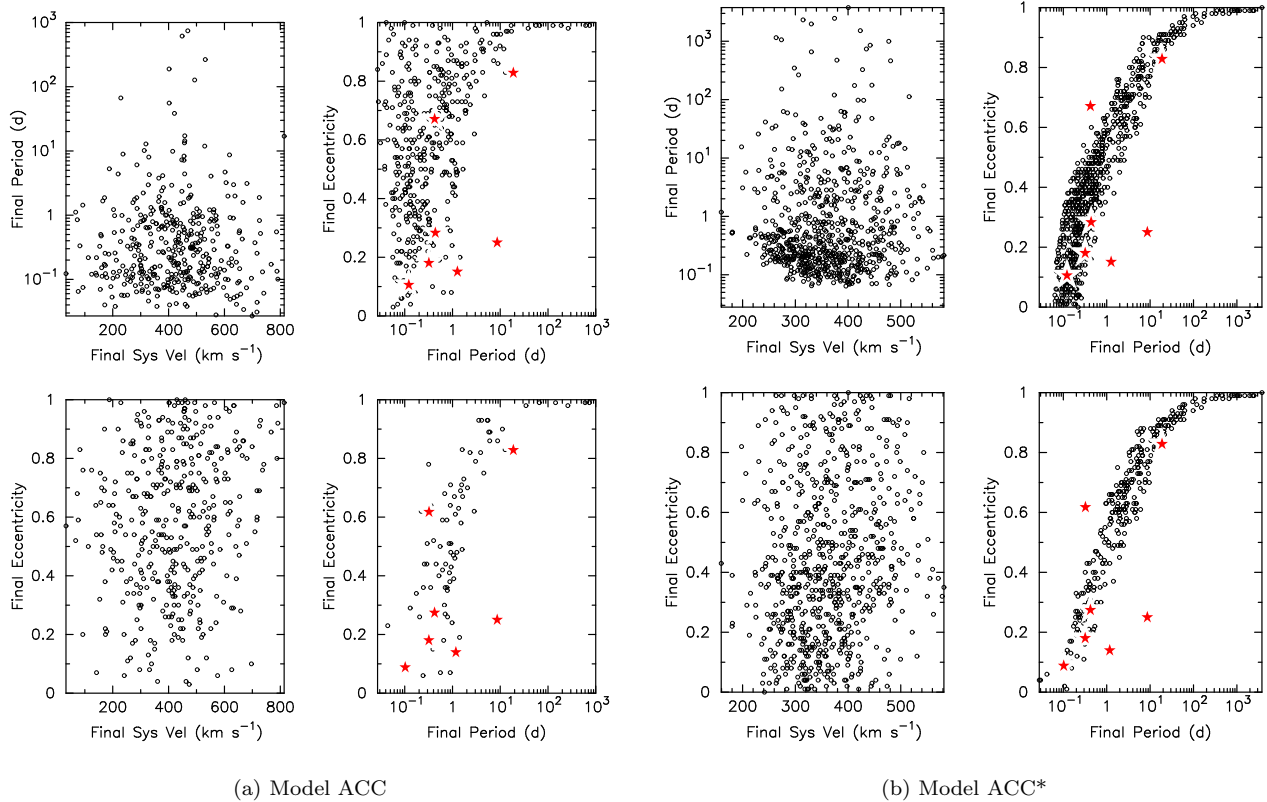

         \vbox{\subfigure[Model
               ACC]{\includegraphics[width=80mm]{dewi2a.ps}}
               \quad \quad \subfigure[Model
               ACC*]{\includegraphics[width=80mm]{dewi2b.ps}}}
         \caption{The distribution of the observable parameters of
         DNSs formed in the double-core scenario, applying a low
         minimum radius for case C mass transfer and $\lambda = 0.1$,
         for model ACC (left) and model ACC* (right). The panels on
         the left of each subfigure show the distributions of the
         orbital period (upper panel) and the eccentricity (lower
         panel) against system velocity. The panels on the right
         present the distributions in the period-eccentricity plane at
         birth (upper) and at present (lower) after taking into
         account the evolution due to gravitational-wave radiation. Also
         plotted as star symbols in the $P-e$ diagrams are the seven
         Galactic DNSs.}
        \label{doublecore:fig:dns}
        \end{figure*}

        \begin{table}
         \caption[]{Birth and merger rates of DNSs, and potential rate
	 of long-duration GRBs for each simulation. Also shown in the
	 table are the disruption rates of DNSs. All rates are in units
	 of $\mathrm{Myr}^{-1}$. The first column indicates the kick
	 velocity distribution (see text for explanation). Asterisks
	 mark models applying low kick velocities to the second-formed
	 neutron star. The second and third columns present the
	 $\lambda$ value and the criterion for the minimum radius for
	 case C mass transfer.}
         \label{doublecore:tab:dns}
         \begin{center}
         \begin{tabular}{lclrrrr}
         Model & $\lambda$ & case C & Birth & Merger &  GRB   & Disruption \\
               &           &        &  rate &  rate  &  rate  &    rate    \\
         \hline
         HP    &       0.1 &  high  &  1.20 &   0.91 &   0.82 &  1.08 \\
         HP    &       0.1 &  low   &  6.92 &   5.86 & 127.67 &  4.74 \\
         HP    &       0.5 &  high  &  0.37 &   0.18 &   0.00 &  1.05 \\
         HP    &       0.5 &  low   & 13.53 &   9.74 &  29.54 & 14.61 \\
         HP*   &       0.1 &  high  &  2.03 &   1.19 &   0.81 &  0.39 \\
         HP*   &       0.1 &  low   & 10.42 &   7.29 & 127.32 &  0.97 \\
         HP*   &       0.5 &  high  &  1.31 &   0.20 &   0.00 &  0.35 \\
         HP*   &       0.5 &  low   & 24.01 &  12.10 &  29.65 &  4.12 \\
         ACC   &       0.1 &  high  &  0.75 &   0.60 &   0.79 &  1.20 \\
         ACC   &       0.1 &  low   &  4.03 &   3.63 & 128.33 &  4.55 \\
         ACC   &       0.5 &  high  &  0.16 &   0.09 &   0.00 &  0.80 \\
         ACC   &       0.5 &  low   &  7.39 &   6.04 &  30.64 & 12.72 \\
         ACC*  &       0.1 &  high  &  1.53 &   1.01 &   0.55 &  0.27 \\
         ACC*  &       0.1 &  low   &  7.34 &   5.43 & 127.28 &  0.54 \\
         ACC*  &       0.5 &  high  &  0.81 &   0.13 &   0.00 &  0.23 \\
         ACC*  &       0.5 &  low   & 17.83 &  10.04 &  29.70 &  2.54 \\
         \end{tabular}
         \end{center}
         \end{table}

When applying a high minimum radius for case C, only systems in relatively wide orbits are allowed to undergo case C mass transfer. Combining the high case-C criterion and a high value of $\lambda$ leads to the formation of post-CE binaries with quite large separations which are much easier to disrupt in the SN explosion. As shown in Table~\ref{doublecore:tab:dns}, this combination yields lower birth and merger rates than applying the high case-C criterion and a low $\lambda$ value.

The picture is rather different when we apply the low case-C criterion.  In this situation, case C mass transfer is allowed also for binaries in much closer orbits. Using low $\lambda$ values results in small separations in such a way that many systems merge. Thus, the combination of the low case-C criterion and low $\lambda$ values results in lower birth and merger rates than applying the low case-C criterion and high $\lambda$ values, as shown in Table~\ref{doublecore:tab:dns}.

Considering the different kick models, model HP yields higher birth and merger rates, by a factor of 1.6 -- 2.3, than model ACC because of the lower dispersion in the kick velocity distribution which leads to more bound systems after the supernova. Similarly, simulations using low kick velocities for the second SN explosion (Models HP* and ACC*) produce higher birth and merger rates. The effect of applying a low kick velocity to the second supernova is more pronounced in models using higher kick velocities for the first-born neutron star, i.e.\ ACC*, in which the rates increase by a factor of 2 -- 5, as opposed to the increase in Model~HP* by as high as a factor of 3.5.

Fig.~\ref{doublecore:fig:dns} presents the results of two simulations of model ACC. The distribution of DNSs in the period-eccentricity plane resembles that obtained from the standard scenario (Dewi, Pols \& van den Heuvel, in preparation -- hereafter DPvdH), i.e.\ there is a main branch in which eccentricity increases with period and an almost vertical branch at low period -- although the two branches here are not as easy to distinguish as in the case of the standard channel. This similarity is due to the fact that the latest stage of evolution in the formation of DNSs is almost identical in the two channels. This indicates that the observed orbital parameters alone cannot be used to distinguish the formation channel through which a DNS was created. Due to the emission of gravitational waves, systems in the vertical branch which are born with low $P$ and high $e$ merge on a relatively short timescale and hence do not appear in the present DNS distribution (bottom-right panel of Fig.~\ref{doublecore:fig:dns}({\it a\/})), as previously discussed by Chaurasia \& Bailes (2005). Simulations using low kick velocities for the second SN do not produce the vertical branch of DNSs with low $P$ and high $e$ such that the present distribution is not significantly different from the distribution at birth (bottom- and upper-left panels of Fig.~\ref{doublecore:fig:dns}({\it b\/})). This situation is also recognized in the standard scenario (DPvdH).

\subsection{Contact phase merger and the progenitor of gamma-ray burst}
\label{doublecore:subsec:grb}

Binary systems that fail to survive the common-envelope phase and merge are potential candidates for long-duration gamma-ray bursts, as the merger of the CO and He cores is likely to produce a rapidly rotating CO star, a plausible candidate for long-duration GRBs in the collapsar model (Woosley 1993; for further recent discussions see Podsiadlowski et al.\ 2004; Fryer \& Heger 2005). The possible GRB rate produced by such mergers is also given in Table~\ref{doublecore:tab:dns}. We obtain the highest GRB rate if we apply the low case-C criterion and a low $\lambda$ value because most of the systems merge during the contact phase. The lowest rate is obtained by using the high case-C criterion and a high $\lambda$ value since practically almost all systems survive the contact phase.

\subsection{Pulsars with black-hole companions}
\label{doublecore:subsec:bh-ns}

We extended our study of the formation of DNSs to include the formation of black-hole/neutron-star (BH-NS) binaries, in which after the contact phase, the CO core collapses and produces a black hole. The initial primary mass is now limited to $8 < M_{1}/{\mathrm{M_{\sun}}} < 40$, but only primary stars with masses $M_{1} > 25$~\msun\ produce black holes (see, e.g., discussions in Pfahl, Podsiadlowski \& Rappaport 2005). We apply two assumptions for the black hole mass; in the first case we assume that the CO core leaves a black hole of 5~\msun, while in the second case we take the black hole mass to be the same as the mass of the CO core progenitor, i.e.\ we asssume that there is no mass loss during black-hole collapse. In the subsequent black-hole/helium-star phase, we assume that the helium star does not experience RLOF because high-mass helium stars generally do not expand significantly (Habets 1986). The orbital evolution is only affected by the wind mass loss from the helium star. Here we apply stellar wind mass loss as in eq.~(2) in Wellstein \& Langer (1999) multiplied by a factor of 0.5. We assume that helium stars less massive than 8~\msun\ become neutron stars, while more massive helium stars leave black holes as the remnants (Pols \& Dewi 2002).

        \begin{table}
         \caption[]{The merger rate of black-hole/neutron-star and
         black-hole/black-hole binaries in units of
         $\mathrm{Myr}^{-1}$. The first column presents the velocity
         dispersion in the kick velocity distribution imparted at the
         collapse of the black holes (the formation of the neutron
         star is always accompanied by a kick velocity with $\sigma =
         290$~\kms). The second and third columns provide the
         $\lambda$ value and the criterion for the minimum radius for
         case C mass transfer. The fourth column presents the assumed
         mass of the black holes.}
         \label{doublecore:tab:bhns}
         \begin{center}
         \begin{tabular}{cclccr}
         $V_{\mathrm{BH}}$ & $\lambda$ & case C & $M_{\mathrm{BH}}$ &
                           BH-NS & BH-BH \\ & & & & rate & rate \\
                           \hline 290 & 0.1 & high & CO & 0.21 & 0.81
                           \\ 290 & 0.1 & high & 5 & 0.14 & 0.63 \\
                           290 & 0.1 & low & CO & 0.89 & 2.57 \\ 290 &
                           0.1 & low & 5 & 0.44 & 2.06 \\ 290 & 0.5 &
                           high & CO & 0.09 & 0.29 \\ 290 & 0.5 & high
                           & 5 & 0.08 & 0.19 \\ 290 & 0.5 & low & CO &
                           1.46 & 5.86 \\ 290 & 0.5 & low & 5 & 1.00 &
                           4.09 \\ 0 & 0.1 & high & CO & 0.85 & 2.38
                           \\ 0 & 0.1 & high & 5 & 0.81 & 2.33 \\ 0 &
                           0.1 & low & CO & 2.19 & 7.20 \\ 0 & 0.1 &
                           low & 5 & 2.26 & 7.07 \\ 0 & 0.5 & high &
                           CO & 0.73 & 2.14 \\ 0 & 0.5 & high & 5 &
                           0.77 & 2.18 \\ 0 & 0.5 & low & CO & 6.32 &
                           19.87 \\ 0 & 0.5 & low & 5 & 6.29 & 19.76
                           \\
         \end{tabular}
         \end{center}
         \end{table}

It is still an open question whether black-hole formation is accompanied by a natal kick imparted to the black hole. The distribution of black-hole X-ray binaries in the Galaxy may indicate that black holes on the whole do not receive significant kicks at birth (White \& van Paradijs 1996). However, the high space velocity of Nova Sco strongly indicates a large natal kick in this particular system (Brandt, Podsiadlowski \& Sigurdsson 1995). The recent study by Jonker \& Nelemans (2004) also supports the idea of a substantial natal kick for the majority of black holes. In this study, we apply two assumptions for the kick velocity imparted to the black hole at birth; one is that black holes are formed without a kick velocity ($\sigma = 0$), and the other that a kick velocity of the same magnitude as in the case of neutron-star formation ($\sigma = 290$~\kms) is imparted.

In the case of almost twin binaries passing through a contact phase, the more massive primary always collapses first with a more massive core. Consequently, in the case of BH-NS binaries, the neutron star is formed later and does not experience any episode of mass transfer, i.e.\ it is not recycled and can only be observable as a young pulsar. Studies on the formation of recycled pulsar with black-hole companion have been carried out by, e.g., Voss \& Tauris (2003), Sipior, Portegies Zwart \& Nelemans (2004), and recently Pfahl et al.\ (2005).

Table~\ref{doublecore:tab:bhns} shows the birth rates of BH-NS and black-hole/black-hole (BH-BH) binaries. All BH-NS and BH-BH systems merge within a Hubble time, and hence the numbers in the table represent both the birth and merger rates. The rates of BH-NS binaries are significantly lower than those of DNSs. The reason is that only a very narrow range of primary masses, i.e.\ $M_1 \sim 25 - 27$~\msun\  can produce BH-NS binaries in the framework of the double-core scenario.  The trend found in DNSs, in which birth and merger rates are the highest for the combination of the low case-C criterion and the high $\lambda$ values, and are the lowest for the high case-C criterion and the high $\lambda$ value, is also applicable to BH-NS binaries. The case where the black-hole mass is equal to its immediate progenitor mass yields higher rates, by typically a factor of 1.5, than the case where mass is lost in the black-hole collapse. However, when we apply zero kicks for black-hole formation, the rates for the two cases of black-hole mass are rather similar.

One would expect to form more BH-NS or BH-BH binaries than NS-BH systems (in which the neutron stars are formed before the black holes) because one requires fine tuning in the initial masses and the orbital period distribution to allow the initially more massive component to have a less massive core at the end of its evolution. The formation of NS-BH binaries is similar to pulsars B2303+46 and J1141--6545, in which white-dwarf companions are formed before the pulsars (Tauris \& Sennels 2000; Davies, King \& Ritter 2002). Indeed, Sipior et al. (2004) found that the birth rate of BH-NS is about 27 higher than NS-BH. This is about 10 times higher than the ratio obtained by Voss \& Tauris (2003).

The birth rates of BH-NS binaries from our simulations are typically comparable or lower than the NS-BH rates found by Voss \& Tauris (2003) or Pfahl et al.\ (2005). The latter authors applied more or less similar assumptions of black-hole masses, kick velocities and $\lambda$ values. The comparable rates between our BH-NS result and their NS-BH binaries is because BH-NS binaries formed by the double-core channel originate from a very narrow mass range, as opposed to the formation of BH-NS through the standard scenario which allows a much wider range of mass to form black hole.

\subsection{Single recycled pulsars}
\label{doublecore:subsec:single}

When the helium star in a HeS-NS binary collapses into a neutron star, it is possible that the system may not survive the SN explosion and become unbound; in this case the two neutron stars become runaway neutron stars. The second neutron star, i.e.\ the descendant of the helium star, will be observed as a young, single radio pulsar, while the first-born neutron star can be observed as a single, recycled pulsar. Pulsars J2235+1506 ($P =$ 55.7~ms; Camilo, Nice \& Taylor 1993) and J0609+2130 ($P =$ 59.7~ms; Lorimer et al. 2004) are two single pulsars with characteristics resembling those of DNSs, and thus have been proposed to have originated from disrupted DNSs. Adopting the commonly accepted high kick velocity at neutron star birth, Lorimer et al.\ (2004) suggested that we should observe more disrupted recycled pulsars than DNSs. Table~\ref{doublecore:tab:dns} displays the disruption rates which, in general, are indeed higher than the birth rates in Models~HP and ACC. The disruption rates are significantly higher than the birth rates in models using the high case-C criterion and high $\lambda$ values because the systems are in relatively wide orbits and hence are more likely to become unbound by the SN. The difference between disruption and birth rates are higher in simulations using ACC kick models due to the large kick velocities.

When we apply a low kick velocity at the second SN, the disruption rates are significantly reduced. The fact that we observe seven DNSs and only two disrupted DNSs is consistent with the suggestion that kick velocities imparted at the birth of the second neutron stars in DNSs are lower than commonly thought.

\begin{figure*}
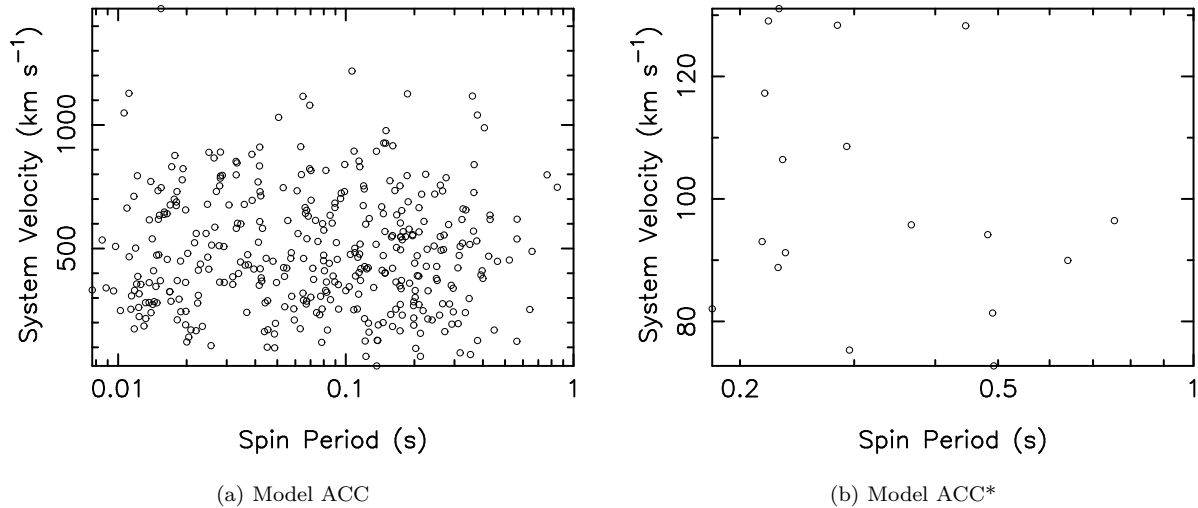

         \vbox{\subfigure[Model
               ACC]{\includegraphics[width=60mm,angle=270]{dewi3a.ps}}
               \quad \quad \subfigure[Model
               ACC*]{\includegraphics[width=60mm,angle=270]{dewi3b.ps}}}
         \caption{The distribution of disrupted recycled single
         pulsars in spin period against space velocity for simulations
         applying low minimum radius for case C mass transfer and
         $\lambda = 0.1$, for Models~ACC and ACC*.}
        \label{doublecore:fig:faildns}
        \end{figure*}

We can derive the runaway velocities of the two neutron stars as in Appendix B in Pfahl, Rappaport \& Podsiadlowski (2002). For the first-born neutron stars, we can calculate the spin period as in Dewi et al.\ (2005) and plot them against the runaway velocities. Fig.~\ref{doublecore:fig:faildns} presents this plot for models~ACC and ACC*. As a consequence of the high kick velocities, disrupted DNSs from Model~ACC have significantly higher runaway velocities. The interesting difference can be seen in the distribution of spin periods; while the spin periods of model~ACC extends to periods as short as 10~ms, only single, recycled pulsars with spin period larger than 0.2~s are produced by model~ACC*. These systems originate from HeS-NS binaries with relatively high-mass helium stars which transfer little amount of mass (hence the large spin period, see the discussion in Dewi et al. (2005)). This rather striking difference may allow us in principle to constrain the kick velocity distribution for the second supernova kick from the spin period distribution of single recycled pulsars from disrupted DNSs.

\section{Discussions and conclusions}
\label{doublecore:sec:conclusion}

We have performed Monte Carlo simulations for the formation of DNSs in the double-core scenario. We found merger rates in this channel of 0.1 -- 12~$\mathrm{Myr}^{-1}$. These rates are subject to substantial uncertainties in the binary input physics. Although it is clear that model~HP provides larger birth and merger rates than model~ACC, we found that the rates are not very sensitive to the choice of kick velocity as both kick models yield rates of the same order of magnitude. The rates are sensitive to the kick velocity only for models HP* and ACC* where we assumed a much lower kick velocity for the
second supernova.

The birth and merger rates depend mainly on the modelling of the contact phase. In this study we limit the investigation to a combination of $\lambda$ values and the minimum radius for case C mass transfer. Although we include primary masses between 8 and 25~\msun, we found that the DNSs mainly originate from primary stars in the range of $\sim$ 12 -- 23~\msun. The $\lambda$ values in this mass range depend sensitively on the stellar mass, but in any case are always substantially lower than 0.5, a standard value used in the previous literature. Our adopted value of $\lambda = 0.1$, even though too low for stars less massive than $\sim$ 14~\msun\ and too high for $M_{1} \gtrsim$\,17~\msun, represents the average $\lambda$ value for stars in the range of $\sim$ 8 -- 25~\msun\ and hence should provide a more reasonable value than $\lambda = 0.5$ -- at least for the primary stars.

Throughout our simulations we apply the same $\lambda$ for the secondary as for the primary. For stars of $\sim$ 16 -- 18~\msun, the $\lambda$ value is relatively constant for a wide range of radius, then increases steeply once the star had finished core helium burning. Since we assume that the primary is in a more advanced stage of evolution and, moreover for this particular study, the primary is at least at the end of the core helium-burning phase, using the same $\lambda$ value for both the primary and secondary is not always the best approach. Estimating the $\lambda$ value for the secondary relative to the value for the primary requires knowledge of the secondary radius at the onset of mass transfer phase and modelling of the early mass-transfer phase, which we were not able to investigate in this study.

Using different minimum radii for case C mass transfer, which simulates the effects of different evolutionary post-main-sequence evolutionary tracks, is one of the major sources of uncertainty for deriving the birth and merger rates of DNSs. Stars less massive than $\sim$ 14~\msun\ generally go through a blue loop and hence the high case-C criterion fits better for primary stars in this range of mass. The situation is not so certain for more massive stars. Knowledge of the primary and secondary radii at the onset of the contact phase is very important for determining the outcome of the contact phase since the radii determine the envelope binding energy (and hence the appropriate value of $\lambda$). It will require detailed binary evolution calculations to overcome the uncertainties on the radii and $\lambda$ values, which we leave for further study.

Even when we know the stellar structure through detailed modelling, we are still left with one important uncertainty, i.e.\ how to model the outcome of a contact phase. In this study we apply the so-called standard prescription in which the final separation is determined by the envelope binding energy.

It is worth noting that our results depend on the initial distributions of the binary parameters, in particular the mass-ratio distribution. If we bias the initial mass-ratio distribution more towards equal masses of both  components, e.g. by simply restricting the mass ratio to a uniform distribution between 0.5 and 1, the resulting rates in Table~\ref{doublecore:tab:dns} are increased by a factor of $\sim 2$.

The double CE scenario studied by Bethe \& Brown (1998) yielded a merger rate of $10^{-5} \, \mathrm{Myr}^{-1}$ for DNSs and $10^{-4} \, \mathrm{Myr}^{-1}$ for BH-NS binaries, which tends to be somewhat higher than most of the rates obtained in this study. We suspect that this huge difference is due to our constraint of only including primaries which go through case C mass transfer. We argue that our additional constraint is required to explain the presence of recycled pulsars.

After discussing the possible uncertainties involved in this study, we will now compare our results with those derived from the standard scenario. DPvdH found birth rates of 0.95 -- 87 $\mathrm{Myr}^{-1}$, comparable and somewhat larger than this study. However, comparison between our results and those of DPvdH is not straightforward as the two studies are done with a very different approach. DPvdH use the BSE binary population synthesis code of Hurley, Tout \& Pols (2002) which, among other factors, includes the loss of angular momentum due to stellar wind mass loss in the orbital evolution. Our double-core study uses a more crude approach. If we also take into account the wind mass loss in our simulations, we suspect that we will have smaller rates. This is because wind mass loss drives the expansion of orbital separation such that some systems may not experience mass transfer phase at all. Wind mass loss depends strongly on stellar luminosity (e.g. de Jager, Nieuwenhuijzen \& van der Hucht 1988; Nugis \& Lamers 2000) and hence for high mass stars investigated in this study is very important. DPvdH also applied actual $\lambda$ value which depends on the evolutionary stage of the star, while we leave it as a constant value.

Nevertheless, DPvdH also found that the birth and merger rates are more sensitive to the modelling of the mass transfer phase preceeding the formation of the HMXBs rather than to the choice of the kick velocity distribution. This suggests that modelling of binary evolution of the stage prior to HeS-NS binaries is very crucial to provide the best merger rates of DNSs.

\section*{Acknowledgements}
JDMD acknowledges a Talent Fellowship from the Netherlands Organization for Scientific Research (NWO) for her stay at the University of Oxford where this work was initiated, and an IoA theory rolling grant from PPARC.

\label{lastpage}

\end{document}